\documentclass
[12pt] {article}
\usepackage{amsmath}
\usepackage{amsfonts}
\usepackage[english]{babel}

\newcommand{\bean}{\begin{eqnarray*}}
\newcommand{\eean}{\end{eqnarray*}}
\newcommand{\ed}{\end{document}}

\newcommand{\be}{\begin{equation}}
\newcommand{\ee}{\end{equation}}
\newcommand{\barr}{\begin{array}}
\newcommand{\earr}{\end{array}}
\newcommand{\bea}{\begin{eqnarray}}
\newcommand{\eea}{\end{eqnarray}}

\newcommand{\proof}{\medskip\noindent{\it Proof.}\quad }
\newcommand{\qed}{\hfill \fbox{}\medskip}

\begin{document}
\title{Classical
BRST
charge and observables in reducible gauge theories} 
\author{Andrei V. Bratchikov 
\\
Kuban State
Technological University,\\Krasnodar, 350072,
Russia
} 
\date {} 
\maketitle

\begin{abstract}
We study the construction of the classical  Becchi-Rouet-Stora-Tyutin (BRST) charge and observables for arbitrary reducible gauge theory. Using a special coordinate system in the extended phase space, we obtain an explicit expression for the Koszul-Tate differential operator and show that the BRST charge can be found by a simple iterative method.
We also give a formula for the classical BRST observables.
\end {abstract}



\section{Introduction}
The modern quantization method for gauge theories is based on the BRST symmetry \cite {BRS},\cite {T}. In the framework of the canonical  formalism this symmetry is generated by the BRST charge. 
If the quantum BRST charge exists it is essentially determined by the corresponding classical one. The classical BRST charge 
is defined 
as a solution to the Poisson-bracket master equation
with certain boundary conditions. 
 The BRST construction in the case of reducible gauge  theories was given in \cite{BF1}. The global existence of the classical BRST charge and observables in the reducible case was proved in \cite {FHST} (see also \cite {HT}). 
In \cite{FHP} the question of quantization of reducible gauge theories with constraints linear in the momenta is studied. 

In this paper we propose another construction of the classical BRST charge for arbitrary reducible gauge theory. To this aim, we 
find a new coordinate system in the extended phase space and transform the master equation by changing variables. This enables us to construct  the Koszul-Tate differential operator. Then the BRST charge can be obtained by using an
iterative method. 
We also give a formula for the classical BRST observables.
As a example of computing the BRST charge we consider an $SU(2)$ gauge invariant reducible theory. 
In this paper we extend the analysis of \cite {B1} to cover the Hamiltonian formalism.

The paper is organized as follows.
In section 2, we review the BRST construction. 
In section 3, we introduce new variables. With respect to the new variables,  the Koszul-Tate differential operator $\delta$ takes a standard form. The construction of the BRST charge is given 
in section~4. 
A formula for the BRST observables is obtained in section 5. In section 6 we find the BRST charge for a reducible theory of order L = 1.

In what follows Grassman parity and ghost number 
of a function $X$ are denoted by $\epsilon (X)$ and $\mbox{gh}(X),$ 
respectively. 
The Poisson superbracket in phase space $\Gamma=(P_A,Q^{A}), \epsilon (P_A)=\epsilon (Q^{A}),$  
is given by 
\begin{eqnarray} \label{sbracket}
\{X, Y\}= \frac {\partial X} {\partial Q^{A}} \frac 
{\partial Y} {\partial P_A} - (-1)^{\epsilon (X)\epsilon (Y)}\frac 
{\partial Y} {\partial Q^{A}} \frac {\partial X} {\partial P_A}.
\end{eqnarray}  
Derivatives with respect to generalized momenta $P_A$ are always understood as left-hand, and those with respect to generalized coordinates $Q^A$ (unless specified) as right-hand ones. The superbracket (\ref {sbracket}) possesses the following algebraic properties
\begin{gather} 
\{X, Y\}= -(-1)^{\epsilon (X)\epsilon (Y)}\{Y,X\},
\phantom{xxxxxxxxxxxxxx\,\,\,\,}
\notag \\
\{X, YZ\}=\{X, Y\}Z +(-1)^{\epsilon (X)\epsilon (Y)}\{Y,X\}Y\{X, Z\},
\notag \\
(-1)^{\epsilon (X)\epsilon (Z)} \{\{X,Y\},Z\}+ \mbox{cycl. perm.}(X,Y,Z)=0.\notag 
\end{gather}  
The last relation is the Jacobi identity for superbracket.
 
\section { 
Generating equations for the gauge algebra
}
Let $P$ be a phase space with the phase space coordinates $\xi_{a},$ $\epsilon(\xi_{a})=\epsilon_{a},$ ${a} =1,\ldots,2m,$ and 
let $G_{a_0},$ ${a_0 =1,\ldots,m_0
,}$ be the 
first class constraints which satisfy the following Poisson brackets
\begin{eqnarray*}
\{G_{a_0},G_{b_0}\}
=U_{{a_0}{b_0}}^{c_0}G_{c_0},
\end{eqnarray*}
where $U_{a_0 b_0}^{c_0}$ are phase space functions. 
The constraints are assumed to be of definite Grassmann parity $\epsilon_{a_0},$ 
$\epsilon(G_{a_0})=\epsilon_{a_0}.$ 

We shall consider a reducible theory of $L$-th order. 
That is, there exist phase space functions 
\begin{eqnarray*}
Z^{a_k}_{a_{k+1}},\qquad  k=0,\ldots,L-1,\qquad a_k=1,\ldots,m_k, 
\end{eqnarray*} 
such that at each stage the $Z$'s form a complete set,
\begin{gather}
Z^{a_k}_{a_{k+1}}\lambda^{a_{k+1}}\approx 0 \Rightarrow \lambda^{a_{k+1}}\approx Z^{a_{k+1}}_{a_{k+2}}\lambda^{a_{k+2}},\qquad k=0,\ldots,L-2, \notag \\
Z^{a_{L-1}}_{a_L}\lambda^{a_L}\approx 0 \Rightarrow \lambda^{a_L}\approx 0.\label {rc} \\
G_{a_0}Z^{a_0}_{a_1}=0,\qquad
Z^{a_{k-2}}_{ a_{k-1}}Z^{a_{k-1}}_{a_{k}}=
V^{a_{k-2}a_0}_{a_{k}}G_{a_0}, \qquad  k=2,\ldots,L.\notag
\end{gather}
The weak equality $\approx$ means equality on the constraint surface 
\begin{eqnarray*}\Sigma:\qquad G_{a_0}=0.
\end{eqnarray*}
Following the BRST method the ghost pairs 
$({\cal P}_{a_k},c^{a_k}),k=0,\ldots,L,$ are introduced 
\begin{align*}
\epsilon({\mathcal P}_{a_k})=\epsilon(c^{a_k})=\epsilon_{a_k}+k+1,\qquad
-\mbox {gh} ({\cal P}_{a_k}) = \mbox {gh} (c^{a_k}) =k+1.
\end{align*}

The BRST charge $\Omega$ is defined as a solution to the equations 
\begin{gather}
\label{1}
\{\Omega,\Omega\}
=0,\\
\label{o7}
\epsilon (\Omega)=1,\qquad
\mbox {gh} (\Omega)=1,
\end{gather}
and satisfying the boundary conditions
\begin{eqnarray*}
\left.
\frac 
{\partial \Omega} 
{\partial c^{a_0}} 
\right|_{c=0} =G_{a_0},\qquad 
\left.\frac {\partial^2 \Omega} {\partial {\mathcal P}_{a_{k-1}}\partial c^{a_k}}\right|_{
{\mathcal P}=c=0} =Z_{a_k}^{a_{k-1}}.
\end{eqnarray*}

One can write
\begin{gather} \label{us}
\Omega= \Omega^{(1)}+M,\qquad M= \sum_{n\geq 2}\Omega^{(n)},\qquad 
\Omega^{(n)}\sim {c}^n ,
\intertext{where}
\Omega^{(1)}= G_{a_0}c^{a_0}+\sum_{k=1}^L \bigl({\mathcal P}_{a_{k-1}}Z_
{a_k}
^{a_{k-1}}+N_{a_k}\bigr)
c^{a_k}
,
\label{sumo}
\end{gather}
$N_{a_1}=0$ and $N_{a_k},$ $k>1,$ only involves ${\mathcal P}_
{a_{s}},s\leq k-2.$ 
Eq. (\ref{o7}) implies 
$\left.N_{a_{k}}\right|_{{\mathcal P}=0}=0 ,
\left. M \right|_{{\mathcal P}=0}=0.$

Let $\cal{B}$ denote 
the algebra of polynomials in 
$({\cal P}_{a_0},c^{a_0},\ldots, {\cal P}_{a_L},c^{a_L})$ with phase space functions coefficients, ${\cal B}=
{\mathbb C}[{\cal P}_{a_0},\ldots, {\cal P}_{a_L}]\otimes 
C^\infty(P)\otimes {\mathbb C}[c^{a_0},\ldots,c^{a_L}].$
Define the subspace 
\begin{eqnarray*}
{\cal U}=\{X\in {\cal B}: X|_{{\cal P}=0,\,\Sigma}=0\}.
\end{eqnarray*}
The space ${\cal U}$ can be decomposed as
${\cal{U}}= \bigoplus_{n\geq 0} {\cal{U}}_{n},$
where ${\cal{U}}_{n}$ is the space of homogeneous polynomials in $(c^{a_0},\ldots,c^{a_L})$ of degree $n.$

\vspace{3mm}
\noindent
\hangindent=2cm \noindent 
{\bf Lemma 1} 
(i) ${\cal U}$ 
is a Poisson subalgebra of 
${\cal B}.$\\
(ii) 
$\Omega 
\in 
{\cal U}.$

\proof
(i) For $X,Y\in {\cal U}$ we have $XY\in {\cal U},$
$\{X,Y\}\in {\cal U}.$ This proves the first statement. 
(ii) 
It is clear that $\Omega^{(1)}\in {\cal U}_1.$
The relation $\Omega^{(n)}\in {\cal U}_n$ for $n\ge 2$ follows from (\ref{o7}) and (\ref{us}). 
\qed 

The bracket $\{.\,,.\}$ splits as 
\begin{eqnarray*} 
\{X,Y\}
=
\{X,Y\}_\xi+ \{X,Y\}_\diamond -(-1)^{\epsilon (X)\epsilon (Y)}\{Y,X\}_\diamond,
\end{eqnarray*}
where $\{.\,,.\}_\xi$ refers to the Poisson bracket in the original phase space and  
\begin{eqnarray*} 
\{X,Y\}_\diamond= \sum_{k=0}^L \frac {\partial X} {\partial c^{a_k}}\frac {\partial Y} {\partial {\mathcal P}_{a_k}} .
\end{eqnarray*}
Let $\delta:{\cal U}\to{\cal U}$ be defined by 
\begin{eqnarray}\label {delta}
\delta =\{\Omega^{(1)},.\}_\diamond= G_{a_0}\frac {\partial  }{\partial {\mathcal P}_{a_{0}}}+\sum_{k=1}^L ({\mathcal P}_{a_{k-1}}Z_{a_k}^{a_{k-1}}+N_{a_k})\frac {\partial } 
{\partial {\mathcal P}_{a_{k}}}.
\end{eqnarray}

Substituting (\ref {us}) in (\ref {1}) one obtains the equations
\begin{align} \label{ru}
\delta \Omega^{(1)}&=0,\\
\label{jjj}
\delta M + D&=0,\end{align}
where
\begin{gather}
D=
\frac 
1 2  F + AM + 
\frac 
1 2 \{M,M\},
\qquad
F=\{\Omega^{(1)}, \Omega^{(1)} \}_\xi,\notag \\
\intertext{
and the operator $A:{\cal U}\to{\cal U}$ is given by}
AX= \{\Omega^{(1)},X \}_\xi-(-1)^{\epsilon(X)}\{X,\Omega^{(1)}\}_\diamond.\notag
\end{gather}
Eq. (\ref{ru}) is equivalent to the nilpotency of $ \delta:$ 
\begin{eqnarray} 
\label{delta2}
\delta^2=0.
\end{eqnarray}

Let ${\cal A}$ denote the Poisson algebra of the first class functions,
\begin{gather}
{\cal A} =\{X(\xi)\,\colon\left. \{X, G_\alpha\}\right|_{\Sigma}=0
\},\notag
\\
\intertext {and let} 
{\cal J} =\{X(\xi)\,\colon \,\left. X\right|_{\Sigma}=0\}. \notag
\end{gather}
Elements of ${\cal A}/{\cal J}$ are called  observables.

A function $\Phi\in {\cal B}
$ is called a BRST-invariant extension of $\Phi_0 \in {\cal A}$ if 
\begin{gather} 
\Phi = \Phi_{0}+\Pi,
\qquad 
\Pi=\sum_{n\geq 1}\Phi_{n},
\qquad 
\Phi_{n} \in {\cal U}_n, \quad n\geq 1,
\qquad 
\mbox{\rm gh} (\Phi)=0,\notag \\
\label {4lll}
\{\Omega,\Phi\}=0.
\end {gather} 
Let $(\mbox {Ker}\,\Omega /\mbox {Im} \Omega)^0$ denote  the set of equivalence classes of BRST-closed functions modulo BRST-exact
functions, with zero ghost number.
The Poisson algebras ${\cal A}/{\cal J}$ and $(\mbox {Ker}\,\Omega /\mbox {Im}\, \Omega)^0$ are isomorphic   
\cite{FHST}.
Elements of $(\mbox {Ker}\,\Omega /\mbox {Im} \Omega)^0$ are called the BRST observables.

\section {Reduction of $\delta$}
In this section we reduce $\delta$ to a standard form.
For $k=L,$ eq. (\ref{rc}) reads 
\begin{eqnarray} 
\label{r1}
Z_{a'_{L-1}}^{a_{L-2}}Z_{a_L^{\phantom {\prime}}}^{a'_{L-1}}+ 
Z^{a_{L-2}}_{a''_{L-1}}Z^{a''_{L-1}}_
{a_L^{\phantom {\prime}}} \approx 0,
\end{eqnarray}
where
$\{a'_{L-1}\},\{a''_{L-1}\}$ are increasing index sets, such that $\{a'_{L-1}\}\cup\, \{a''_{L-1}\}= \{a_{L-1}\},$ $|\{a'_{L-1}\}|=|\{a_L\}|$ and ${\rm rank}\,  Z^{a'_{L-1}}_{a_L^{\phantom{\prime}}}=|\{a_L\}|.$ 
For an index set $i= \{i_1,i_2,\ldots,i_n\},$ we denote $|i|=n.$
 From  (\ref{r1}) it follows
that ${\rm rank}\, Z^{a_{L-2}}_{a_{L-1}}=|\{a_{L-1}\}|-|\{a_L\}|=|\{a''_{L-1}\}|,$ and  ${\rm rank}\, Z^{a_{L-2}}_{a''_{L-1}}=| \{a''_{L-1}\}|.$

One can split the index set $\{a_{L-2}\}$ as
$\{a_{L-2}\}=\{a'_{L-2}\}\cup\, \{a''_{L-2}\},$ such that  ${|\{a'_{L-2}\}|=|\{a''_{L-1}\}|,}$ and ${\rm rank}\, Z^{a'_{L-2}}_{a''_{L-1}}=|\{a''_{L-1}\}|.$
For $k=L-1,$ eq. (\ref{rc}) implies 
\begin{eqnarray*} 
\label{r2}
Z_{a'_{L-2}}^{a_{L-3}}Z_{a''_{L-1}}^{a'_{L-2}}+Z^{a_{L-3}}_{a''_{L-2}}
Z^{a''_{L-2}}_{a''_{L-1}}\approx 0.
\end{eqnarray*}
From this it follows that 
\begin{eqnarray*}
{\rm rank}\, Z^{a_{L-3}}_{a''_{L-2}}=
{\rm rank}\, Z^{a_{L-3}}_{a_{L-2}}=|\{a_{L-2}\}|-|\{a''_{L-1}\}|=|\{a''_{L-2}\}|.
\end{eqnarray*}

Using induction on $k,$ we obtain a set of nonsingular matrices $Z^{a'_{k-1}}_{a''_{k}},$
$k=1,\ldots,L,$
and a set of matrices $Z^{a_{k-1}}_{a''_k},k=1,\ldots,L,$ such that
\begin{eqnarray*}
{\rm rank}\, Z^{a_{k-1}}_{a''_{k}}= {\rm rank}\, Z^{a_{k-1}}_{a_{k}}=|\{a''_k\}|.
\end{eqnarray*}
Here $\{ a'_k\}\cup \{a''_k\}=\{a_k\},$
$k=1,\ldots,L-1,$ 

Eq. (\ref{rc}) implies 
\begin{eqnarray} \label{o2}
G_{a'_0}Z^{a'_0}_{a''_1} + G_{a''_0} Z^{a''_0}_{a''_1}= 0.
\end{eqnarray}
From this it follows that $G_{a''_0}$ are independent. 
We assume that $G_{a''_0}$ satisfy the regularity conditions. It means that there are some functions $F_{\alpha}(\xi),$ $\{\alpha\}\cup \{a''_0\}=\{a\},$ such that $(F_{\alpha},G_{a''_0})$  can be locally taken as new coordinates in the original phase space.

Let $f: \{a''_{k+1}\} \to \{a_k\},$ $k=~0,\ldots, L-~1,$ be an embedding, $ f(j)=j,$
and let $\{\alpha_k\}$ be defined by $\{a_k\}= \{f(a''_{k+1})\}\cup \{\alpha_k\}.$ Since $|\{a''_k\}|=|\{\alpha_k\}|,$ one can write $\alpha_k=g(a''_{k})$ for some function $g,$ and hence
$$\{a_k\}= \{f(a''_{k+1})\}\cup \{g(a''_k)\}, \qquad k=~0,\ldots, L-~1.$$

\vspace{3mm}
\noindent
{\bf Lemma 2} 
{\it The nilpotent operator $\delta$ is reducible to the form 
\begin{eqnarray} 
\label{de9}
\delta= \xi'_{a''_0} \frac{\partial}{\partial {\mathcal P}'_{g(a''_0)}}
+\sum_{k=1}^{L} {\mathcal P}'_{f(a''_k)}\frac{\partial }{\partial {\mathcal P}'_{g(a''_k)}},
\end{eqnarray}
by the change of variables:
$(\xi_a,{\mathcal P}_{a_0}, \ldots, {\mathcal P}_{a_{L}})\to
(\xi'_a,{\mathcal P}'_{a_0}, \ldots, {\mathcal P}'_{a_{L}}),$
\begin{gather} 
\xi'_{\alpha} = F_{\alpha},\qquad \xi'_{a''_0} =  G_{a''_0},\notag \\
\label {cha} 
{\mathcal P}'_{
f(a''_{k+1})}=\delta{\mathcal P}_{a''_{k+1}}
,\qquad {\mathcal P}'_{g(a''_{k})}={\mathcal P}_{a''_{k}},\\
{\mathcal P}'_{a_{L}} = {\mathcal P}_{a_{L}},\notag
\end{gather}
where $k=0,\ldots,L-1,$ $g(a''_L)=a_L.$
}

\proof 
To prove this statement we first observe that eqs. \eqref {cha} are  solvable with respect to $(\xi_a,{\mathcal P}_{a_0}, \ldots, {\mathcal P}_{a_{L}}).$ 
The original variables can be represented as  
\begin{eqnarray*}\label{cucu}
\xi_{a}= \xi_{a}(\xi'),\qquad {\mathcal P}_{a_k}={\mathcal P}_{a_k}(\xi'_{a},
{\mathcal P}'_{a_0},\ldots, {\mathcal P}'_{a_k}),\qquad  k=0,\ldots,L.
\end{eqnarray*}
Here we have used the fact that the ${\mathcal P}_{a_k}$ depends 
only on the functions ${\mathcal P}'_{a_s}$ with $s\leq k.$
Assume that the functions $\xi_{a}(\xi')$ have been constructed.
Then from (\ref{cha}) it follows that
\begin{align}
{\mathcal P}_{a'_k}&=
({\mathcal P}'_{f(a''_{k+1})}- 
{\mathcal P}'_{g(a''_k)} Z^{\prime a''_{k}}_{a''_{k+1}}- N^{\prime\phantom{a_k}}_{a''_{k+1}})(Z^{\prime (-1)})^{a''_{k+1}}_{a'_k},\notag \\
\label{cucu22}
{\mathcal P}_{a''_k}&={\mathcal P}'_{g(a''_k)},\qquad k=0,\ldots,L-1,
\\
{\mathcal P}_{a_{L}}
&
= {\mathcal P}'_{a_{L}}.
\notag
\end{align}
Here and in what follows
\begin{eqnarray*}
X'(\xi', {\mathcal P}'_{a_0},\ldots, {\mathcal P}'_{a_{L}})=
X(\xi, {\mathcal P}_{a_0},\ldots,{\mathcal P}_{a_{L}}).
\end{eqnarray*}
Using (\ref {delta}) and (\ref {delta2}) one gets 
\begin{gather}
\delta \xi'_a=\delta {\mathcal P}'_{f(a''_1)}=\ldots=\delta {\mathcal P}'_{f(a''_L)}=0,\notag \\
\label{cu55}
\delta {\mathcal P}'_{g(a''_0)}=\xi'_{a''_0}, \qquad \delta {\mathcal P}'_{g(a''_k)}= {\mathcal P}'_{f(a''_{k})}, \qquad k=1,\ldots, L.
\end{gather}
 Eqs. (\ref {cu55}) are equivalent to (\ref {de9}).
\qed

With respect to the new coordinate system the condition $X\in {\cal U}$ 
implies $$\left. X \right|_{\xi'_{a''_0}={\mathcal P}'=0}=0.$$

\section {The BRST charge} 
\paragraph {Constructing of $\delta.$
}
To construct $\delta$ we need 
the functions $N_{a_2},\ldots,N_{a_L},$
which are defined by the following system of recurrent equations 
\begin{eqnarray} \label {f}
\delta N_{a_k}=-({\cal P}_
{a_{k-2}}Z_{a_{k-1}}^{a_{k-2}}+
N_{a_{k-1}})Z_{a_k}^{a_{k-1}},\qquad k=2,\ldots,L, 
\end{eqnarray}
with $N_{a_1}=0,$
$N_{a_k}\in {\cal V}_{k-2},$ where ${\cal V}_{k-2}$
is the subspace of ${\cal U}$ which consists of
the functions depending only on $(\xi_{a},{\cal P}_{a_0}, \ldots, {\cal P}_{a_{k-2}}).$ 
Let $\delta_{k}$ denote the restriction of $\delta$ on ${\cal V}_{k-2},$
$$\delta_{k}= \xi'_{a''_0} \frac{\delta }{\delta{\cal P}'_{g(a''_0)}}
+\sum_{s=1}^{k-2} {\cal P}'_{f(a''_s)}\frac{\delta }{\delta {\cal P}'_{g(a''_s)}}, 
$$
let $n_{k}$ 
be the counting operator 
$$
n_{k} = \xi'_{a''_{0}}\frac{\delta_l }{\delta \xi'_{a''_{0}}}+ {\cal P}'_{g(a''_{0})}\frac{\delta}{\delta 
{\cal P}'_{g(a''_{0})}}+ \sum_{s=1}^{k-2}\left({\cal P}'_{f(a''_{s})}\frac{\delta}{\delta {\cal P}'_{f(a''_{s})}}
+  {\cal P}'_{g(a''_{s})}\frac {\delta} {\delta {\cal P}'_{g(a''_{s})}}\right),
$$
and let 
$$ 
\sigma_{k}= {\cal P}'_{g(a''_0)} \frac{\delta_l}{\delta \xi'_{a''_0}}+
\sum_{s=1}^{k-2} {\cal P}'_{g(a''_{s})}\frac{\delta }{\delta {\cal P}'_{f(a''_{s})}}
.
$$
One can directly verify that 
\begin{eqnarray} \label{us4} 
\delta^2_k=\sigma^2_{k}=0,\qquad \delta_{k}\sigma_{k}+\sigma_{k} \delta_{k}=n_{k},\qquad n_{k}\delta_{k}=\delta_{k} n_{k} , 
\qquad n_{k}\sigma_{k}=\sigma_{k} n_{k}.
\end{eqnarray} 

The space ${\cal V}_{k-2}$ splits as 
$$ 
{\cal V}_{k-2}= {\cal V}_{k-2}^{(0)}\oplus\widetilde{\cal V}_{k-2}, \qquad 
\widetilde{\cal V}_{k-2}= {\cal V}_{k-2}^{(1)} \oplus{\cal V}_{k-2}^{(2)} \oplus \ldots,
$$
with  $
n_k X=nX$ for $X\in {\cal V}_{k-2}^{(n)}.$ 
It is clear that  
\begin {align}
{\cal V}_{k-2}^{(0)}&=\{ \Phi\in {\cal V}_{k-2}\,|\, \Phi=\Phi( \phi'_{a'_0},
\phi^{*\prime}_{f(a''_{k-1})})
\},\qquad k\ne L+2,\notag \\ 
\label{xxxx} {\cal V}_{L}^{(0)}&=0.
\end {align}

The subspace $\widetilde {\cal V}_{k-2}$ is invariant under the action of 
$\delta_{k},$ $\sigma_{k}$ and $n_k.$
The operator $n_k: \widetilde {\cal V}_{k-2} \to \widetilde {\cal V}_{k-2}$ is invertible. 
It follows from (\ref{us4}) that   $\delta_{k}^{{+}}: \widetilde {\cal V}_{k-2}\to \widetilde {\cal V}_{k-2},$ defined by $\delta_{k}^{{+}}=\sigma_{k} n_{k}^{-1},$ is a generalized inverse of $\delta_{k}$: 
\begin{eqnarray} \label{mk} 
\delta_{k}\delta_{k}^{+}\delta_{k}=\delta_{k},
\qquad \delta_{k}^{+}\delta_{k}\delta_{k}^{+}=\delta_{k}^{+}
,
\end{eqnarray} 
and for any $X
\in \widetilde{\cal V}_{k-2},$  
\begin {eqnarray}
\label{osas}
X=\delta^+_k\delta_k X+\delta_k\delta^+_kX.
\end {eqnarray}

\vspace{3mm}
\noindent

{\bf Theorem 1.} {\it The general solution to (\ref{f}) is given by
\begin{eqnarray} \label {kkk}
 N_{a_k}=Y_k-\delta^+_k\left(({\cal P}_
{a_{k-2}}Z_{a_{k-1}}^{a_{k-2}}+
N_{a_{k-1}})Z_{a_k}^{a_{k-1}}\right),\qquad k=2,\ldots,L, 
\end{eqnarray}
where $Y_k$ 
is an arbitrary cocycle, 
$\delta Y_k=0,$ 
subject only to the restrictions 
\begin{eqnarray} 
\label{eee}
Y_{a_k}\in \widetilde {\cal V}_{k-2} ,\qquad
 \epsilon(Y_{a_k})=\epsilon(N_{a_k}),\qquad
\mbox{\rm gh}(Y_{a_k})=\mbox{\rm gh}(N_{a_k}).
\end{eqnarray}
}

{\it Proof.} Assume that the functions $N_{a_{s}}\in  \widetilde {\cal V}_{s-2},$ $s<k,$ 
have been constructed.
Changing variables in \eqref {f} $
(\xi^{a},{\cal P}_{a_0},\ldots, {\cal P}_{a_{k-2}})\to
(\xi'_{a},{\cal P}'_{a_0},\ldots, {\cal P}'_{a_{k-2}}),
$
we get 
\begin{eqnarray} \label{f1}
\delta_{k}  N'_{a_k}= D'_{a_k},
\end{eqnarray}
where $$D'_{a_k}= - ({\cal P}_
{a_{k-2}}Z_{a_{k-1}}^{\prime a_{k-2}}+ N'_{a_{k-1}})Z_{a_k}^{\prime a_{k-1}},
\qquad
{\cal P}_
{a_{k-2}}= {\cal P}_{a_{k-2}}(\xi'
,{\cal P}'
).
$$
Eq. (\ref{rc}) reads
\begin{eqnarray*} 
Z^{\prime a_{k-2}}_{ a_{k-1}}Z^{\prime a_{k-1}}_{a_{k}}=V^{\prime a_{k-1}a'_0}_{a_{k}}
G'_{a'_0}
+
V^{\prime a_{k-1}a''_0}_{a_{k}}
\xi'_{a''_0}.
\end{eqnarray*}
It follows from (\ref{o2}) that 
\begin{eqnarray*} 
G'_{a'_0}=-\xi'_{a''_0} Z^{\prime  a''_{0}}_{ a''_{1}}(Z^{\prime (-1)}) ^{a''_{1}}_{a'_{0}}
,
\end{eqnarray*}
and hence
\begin{eqnarray*}
\label {rc2}  
Z^{\prime a_{k-2}}_{ a_{k-1}}Z^{\prime a_{k-1}}_{a_{k}}\in 
\widetilde {\cal V}_{k-2}.
\end{eqnarray*}
Therefore 
$D'_{a_k}\in \widetilde {\cal V}_{k-2}.$

One can directly verify that $\delta_k  D'_{a_k}=0,$ or equivalently,  using (\ref{osas}),
$\delta_{k}\delta_{k}^{+}D'_{a_k}=D'_{a_k}.$
Then the general solution to (\ref {f1}) 
is given by
\begin{eqnarray}
 \label{o10}
N'_{a_k} = Y'_{a_k} + \delta_{k}^{{+}}D'_{a_k},
\end{eqnarray} 
where the cocycle
$Y'_{a_k}$ satisfies (\ref{eee}). 
By its construction, $N'_{a_k}\in \widetilde {\cal V}_{k-2}.$ In the original variables (\ref{o10}) takes the form (\ref {kkk}).
\qed

\paragraph {Higher orders.}
Our next task is to find a solution to  
(\ref {jjj}).  
We shall need the following Lemma.

\vspace{3mm}
\noindent
{\bf Lemma 3} {\it Let $R$
denote the left-hand side of (\ref {jjj}), 
\begin{eqnarray} \label {www}
R=\delta M +D.
\end{eqnarray}
 Then
\begin {eqnarray}
\label{omf}
\delta R+ AR + \{M, R\}=0. 
\end {eqnarray}}

\proof
 If 
(\ref{delta2}) holds, then $R=\{\Omega,\Omega\}.$ 
From the Jacobi identity ${\{\Omega,\{\Omega,\Omega\}\}=0}$ it follows that $\{\Omega,R\}=0,$ which is equivalent to (\ref{omf}).
Here we have used the relation 
$
\{\Omega^{(1)},\,.\,\}=\delta+A.$
\qed

Since $ {\cal U}=  
{\cal V}_L\otimes {\mathbb C}[c^{a_0},\ldots,c^{a_L}]$, it follows from (\ref{xxxx}) that
the operator $n_{L+2}\colon {\cal U} \to {\cal U}$ is invertible. Denote $\delta^+=\delta^+
_{L+2}.$
The space ${\cal U}$ can be decomposed as
\begin{eqnarray} 
\label{o25}
{\cal U}=
\mbox {Ker}\, \delta
\oplus 
\mbox {Ker}\, \delta^+,
\end{eqnarray} 
where the corresponding orthogonal projectors are given by
\bean
P_{Ker\, \delta}
=\delta\delta^+,\qquad 
P_{Ker\, \delta^+}
=I- \delta\delta^+=\delta^+\delta,  
\eean
 $I$ is the identity map.The last relation follows from (\ref{us4}).

Let $\langle .\,,. \rangle: {{\cal U}}^2\to {{\cal U}}  $ be defined by
\begin{eqnarray} \label {metric}
\langle X,Y \rangle = -\frac 1 2 (I+\delta^+A)^{-1}\delta^{+}\left(\{ X,Y\}+\{Y,X\}\right).
\end{eqnarray} 
where
$(I+\delta^+A)^{(-1)}=\sum_{m\geq 0}(-1)^m(\delta^+A)^m.$

\vspace{3mm}
\noindent

{\bf Theorem 2} {\it The general solution to 
(\ref {jjj}) can be obtained by applying the iteration method  to the equation
\bea 
\label{ome}
M = M_0 + \frac 1 2 \langle M,M \rangle,
\eea
where 
\bea 
\label{mo}
M_0 = (I+\delta^+A)^{-1}\left({W}-\frac 1 2 \delta^{+}F\right),
\eea 
$W$ is an arbitrary 
cocycle, 
$\delta W=0,$ 
subject only to the restrictions 
\begin{eqnarray} \label{ccc}
\epsilon(W)=1,\qquad \mbox{\rm gh}(W)=1,\qquad
W\in \bigoplus
_{n\ge 2}{\cal U}_n.
\end{eqnarray}
}

\proof 
In accordance with the decomposition (\ref{o25})
eq. (\ref {jjj}) splits as  
\begin{align} \label{c}
\delta M +\delta\delta^+ D&=0,\\
\label {ddd}
(I-\delta\delta^+) R&=0.
\end{align}

From (\ref{c}) it follows that 
\begin{eqnarray} M + \delta^+ D= W ,
\label {mu}
\end{eqnarray}
where the cocycle $W$ satisfies (\ref {ccc}). 
Using (\ref {metric}), one can write (\ref{mu}) in the form (\ref {ome}). Eq. (\ref {ome}) can be iteratively solved as:
\begin{eqnarray}
\label{omey}
M = M_0 + \frac 1 2 \langle M_0,M_0 \rangle +\ldots.
\eea

To prove that the solution to (\ref {mu}) satisfies (\ref {ddd}) we use the approach of ref. \cite{F}.
 Consider (\ref{omf})
and the condition
\begin {eqnarray}
\label{ppp}
\delta^+R=0. 
\end {eqnarray}
Applying $\delta^+$ to (\ref{omf}) and using (\ref{ppp}), we get 
\begin {eqnarray}
\label{yyy}
 R=-\delta^+ (A R + \{M,R\}). 
\end {eqnarray}
From (\ref{yyy}) by iterations it follows that $R=0.$ 

It remains to check \eqref {ppp}. The solution to (\ref{mu}) satisfies  
$\delta^+M = \delta^+W,$ which implies 
\begin {eqnarray}
\label{osm}
 M=\delta^+\delta M + W. 
\end {eqnarray}
By definition (\ref{www}), we have 
$\delta^+R=\delta^+\delta M+\delta^+D,$
and therefore by (\ref{mu}) and (\ref{osm}), $\delta^+R=0.$ 
\qed

Series (\ref{omey}) can be obtained by using a diagram  technique \cite{B2}. 

\section {The BRST observables} 
Consider the equation 
\begin {eqnarray}  
\label {linm}
\{\Omega,\Psi\}-\Lambda=0,
\end {eqnarray} 
where $\Lambda$ is a given function, $\Lambda\in {\cal U},$ $\epsilon(\Lambda)=1,$ ${\rm gh}(\Lambda)=1,$ $\{\Omega,\Lambda\}=0,$ and $\Psi\in {\cal U}$ is an unknown one.
 
\vspace{3mm}
\noindent
{\bf Lemma 4} \cite {BT}  {\it Any solution to the homogeneous equation  \begin {eqnarray*}  
\label {nnn}
\{\Omega,\Psi\}=0,\qquad \Psi\in {\cal U},
\end {eqnarray*}  
is given by
$$\Psi=\{\Omega,\Upsilon\},$$
where
\begin {eqnarray}\label {nnn}\Upsilon\in {\cal U},\qquad
\epsilon (\Upsilon)=1,\qquad {\mbox {gh} }(\Upsilon)=-1.\end {eqnarray}} 




Denote $\{X,\,.\,\}$ by $ {\rm ad}\,X.$

\vspace{3mm}
\noindent
{\bf Theorem 3} {\it The general solution to (\ref {linm}) is given by 
\begin {eqnarray}  
\label {linm3}
\Psi=\Psi_p+\{\Omega,\Upsilon\},
\end {eqnarray}
where
\begin {eqnarray} 
\label{qppp}
\Psi_p=(I+\delta^{+}(A+ {\rm ad}\,M) )^{-1
}\,\delta^{+}\Lambda,
\end {eqnarray}  
and $\Upsilon$ satisfies (\ref{nnn}). }

\proof 
Eq. (\ref{linm}) can be written as
\begin {eqnarray}
\label{qspm}
\delta \Psi+ Q=0,
\end {eqnarray}
where 
$Q=A \Psi + \{M, \Psi\}- \Lambda.$ 
Denote by $\Gamma$ the left-hand side of (\ref{linm}),
\begin {eqnarray}  
\label {222}
{\Gamma=\{\Omega,\Psi\}-\Lambda}=\delta \Psi+ Q.
\end {eqnarray} 
Using (\ref{o25}), we split (\ref {qspm}) into the following two equations
\begin {align} 
\label{qqq}
\delta  \Psi+ \delta \delta^+ Q&=0,\\
\label{mmm}
(I-\delta \delta^+)\Gamma&=0.
\end {align}
Eq. (\ref{qqq}) is equivalent to 
\begin {eqnarray}
\label{sss}
\Psi+ \delta^{+}Q=\Upsilon,
\end {eqnarray}
where $\Upsilon$ is a cocycle,  $\delta \Upsilon=0.$ Setting $\Upsilon=0,$ we 
get from (\ref{sss}) the particular solution $\Psi_p$ (\ref{qppp}).

Now, let us show that (\ref{qppp}) satisfies (\ref{mmm}). From the Jacoby identity $\{\Omega,\{\Omega,\Psi\}\}=0$ and the BRST invariance of $\Lambda$ it follows that 
\begin {eqnarray}
\label{xxx}
\delta \Gamma+ A\Gamma + \{M, \Gamma\}=0. 
\end {eqnarray}

It is straightforward to check that
\begin {eqnarray}
\label{ttt}
\delta^{+}\Gamma=0.
\end {eqnarray}
Indeed, (\ref {222}) implies that
\begin {eqnarray*}
\delta^{+}\Gamma=\delta^{+}(\delta \Psi+ Q)=\delta^{+}\delta \Psi- \Psi=0,
\end {eqnarray*}
which gives (\ref{ttt}) since
$\Psi\in \mbox {Ker}\, \delta^+\cap {\cal U}.$  
Applying $\delta^+$ to (\ref{xxx}) and using  (\ref{ttt}) we get  
\begin {eqnarray}
\Gamma=-\delta^{+}(A\Gamma + \{M, \Gamma\}), 
\end {eqnarray} from which it follows that $\Gamma=0.$ 
We conclude that (\ref{mmm}) is satisfied by $\Psi_p$ (\ref {qppp}). Using Lemma 4 we get 
(\ref{linm3}).
 \qed

\vspace{3mm}
\noindent
{\bf Corollary} {\it The BRST invariant extention of $\Phi_0$ is given by  
\begin {eqnarray} 
\label{q221}
\Phi=\Phi_0-(I+\delta^{+}(A+ {\rm ad}\,M) )^{-1}\delta^{+}\{\Omega,\Phi_0\}+\{\Omega,\Upsilon\}
\end {eqnarray} 
with some function $\Upsilon$ satisfying  (\ref{nnn}). }

\proof Eq. (\ref{4lll}) reads 
$\{\Omega,\Pi \}+\Delta=0,$
where $\Delta=\{\Omega,\Phi_0\}.$ By construction, $\Pi,\Delta\in {\cal U},$ 
$\epsilon(\Delta)=1,$ ${\rm gh}(\Delta)=1.$
From the Jacoby identity $\{\Omega,\{\Omega,\Phi_0\}\}=0$ it follows that 
$\{\Omega,\Delta\}=0.$ Eq. (\ref{q221}) is obtained by applying Theorem 3. 
\qed

Since ${\rm ad}\,\Omega=\delta+A+ {\rm ad}\,M,$ (\ref {q221}) can be rewritten as 
\begin {eqnarray*} 
\Phi=\Phi_0-(I+\delta^{+}({\rm ad}\,\Omega-\delta) )^{-1}\delta^{+}\{\Omega,\Phi_0\}+\{\Omega,\Upsilon\}.
\end {eqnarray*} 
Using (\ref{q221}) 
we can effectively construct elements of  $(\mbox {Ker}\,\Omega /\mbox {Im} \,\Omega)^0$ for arbitrary reducible gauge theory.

\section {$SU(2)$ gauge invariant reducible theory of order $L = 1.$
} 
To illustrate the method of 
computing the BRST charge 
let us consider a simple reducible model. 
The model
is described by three pairs of canonically conjugate
variables $(\varphi_a, \pi_a).$ It is subject to the first class constraints
\begin{eqnarray}\label{co33} 
G_a=\varepsilon_{abc}\varphi_b\pi_c.
\end{eqnarray}
The algebra
of these functions is the $su(2)$ Lie algebra 
:\begin{eqnarray*} 
\{G_a,G_b\} = \varepsilon_{abc}G_c.
\end{eqnarray*}
Constraints
(\ref{co33}) 
appear 
in the Yang-Mills quantum mechanics \cite{Sh}. 
The reducibility condition reads
\begin{eqnarray*} 
G_a\pi_a =0.
\end{eqnarray*}

$\Omega^{(1)},$ $F$ and  $\delta$ are given by 
\begin{gather*}
\Omega^{(1)}= G_{a}c^{a}+{\cal P}_{a} \pi_{a}c,\qquad
F=G_a\varepsilon_{abc}c^bc^c
-2{\cal P}_a\varepsilon_{abc}\pi_bc^cc,\\
\delta=G_a
\frac {\partial} 
{\partial {\cal P}_a}+
{\cal P}_a\pi_a\frac {\partial } 
{\partial {\cal P}},
\end{gather*}
where $(c^a,{\cal P}_a)$ and $(c,{\cal P})$ are auxiliary canonically conjugated
variables,
\begin{gather*}
\epsilon({c}^a)=\epsilon({\cal P}_a)=1,\qquad
\epsilon({c})=\epsilon({\cal P})=0,\qquad
\mbox{gh}({c}^a)=-\mbox{gh}({\cal P}_a)=1,\\
\mbox{gh}({c})=-\mbox{gh}({\cal P})=2.
\end{gather*}


The change of variables
\begin{eqnarray} 
\label{olya}
\pi'_{i}=G_{i},\qquad \pi'_{3} = \pi_{3},\qquad
{\cal P}'_{i} = {\cal P}_{i},\qquad{\cal P}'_{3} ={\cal P}_{a}\pi_a
,\qquad {\cal P}' = {\cal P},
\end{eqnarray}
where $ i=1,2,$ yields
\begin{gather*} 
\delta= 
\pi'_i \frac {\partial}{\partial {\cal P}'_i}+
{\cal P}'_3 \frac{\partial}{\partial {\cal P}'}
,\qquad \sigma= {\cal P}'_{i} \frac{\partial}{\partial \pi'_i}
+{\cal P}'\frac{\partial }{\partial {\cal P}'_3},\\
n = \pi'_i
\frac {\partial }{\partial \pi'_{i}}+ {\cal P}'_i\frac{\partial }{\partial {\cal P}'_i}+
{\cal P}'\frac{\partial}{\partial {\cal P}'}.
\end{gather*}
 In the domain with ${\varphi_3}\ne 0,$ ${\pi_3}\ne 0,$ 
the transformation
(\ref{olya}) is invertible: 
\begin{eqnarray*}
{\pi}_{i}=\frac 1 {\varphi_3}
\left (\varepsilon_{ij} {\pi}'_j+{\varphi_i}\pi'_3\right)
,\qquad  
{\pi}_3= {\pi'_3},
\qquad
{\cal P}_{i}=
{\cal P}'_i,
\qquad  
{\cal P}_3=\frac 1 {\pi'_3}\left({\cal P}'_3-
{\cal P}'_{i}\pi_i\right).
\end{eqnarray*}
Here $\varepsilon_{ij}=\varepsilon_{ij3},$
$\pi_i=\pi_i(\pi').$

One gets
\begin{align*}
\delta^+F' &= \left({\cal P}'_i\varepsilon_{ibc}-\frac 1 {\varphi_3}{\cal P}'_i\varphi_i\varepsilon_{3bc}\right)c^bc^c
- \\
 & -\frac 2 {\varphi_3}
\left(
\frac 1 {\pi'_3}\left(\varepsilon_{ij}{\cal P}'_i{\cal P}'_j+{\varphi_3}{\cal P}'
\right)
\varepsilon_{ij}\pi_i(\pi')-
{{\cal P}_3 (\pi',{\cal P}')}
{\cal P}'_j
\right)c^jc.
\end{align*}
To obtain a regular expression for $\Omega$
we take 
\begin{eqnarray*} \label {or}
W'=-\frac 1 {\pi'_3}\left({\cal P}'_3 +
\frac 1\varphi_3\varepsilon_{ij}\pi'_i {\cal P}'_j \right)\varepsilon_{kl}c^kc^l.
\end{eqnarray*} 
Then, one finds
\begin{eqnarray}\label{bbb4}
W'-\frac 1 2  \delta^+F'=-\frac 1 2(I+\delta^+A) {\cal P}_{a} \varepsilon_{abc}c^{b}c^{c}.
\end{eqnarray}
Substitution (\ref{bbb4}) in (\ref{mo}) yields
\begin{eqnarray*}
M_{0}= -\frac 1 2 {\cal P}_{a} \varepsilon_{abc}c^{b}c^{c}.
\end{eqnarray*}
Since $\{ {\cal P}_{a} \varepsilon_{abc}c^{b}c^{c},{\cal P}_{d} \varepsilon_{def}c^{e}c^{f}\}=0,$ it follows from (\ref{omey}) that $M=M_0,$ and hence
\begin{eqnarray*}
\Omega= G_{a}c^{a}+{\cal P}_{a} \pi_{a}c-\frac 1 2 {\cal P}_{a} \varepsilon_{abc}c^{b}c^{c}.
\end{eqnarray*}
 

\end{document}